\documentstyle[12pt,epsfig,epsf]{article}
\def\beq{\begin{equation}}
\def\eeq{\end{equation}}
\def\bea{\begin{eqnarray}}
\def\eea{\end{eqnarray}}
\def\bq{\begin{quote}}
\def\eq{\end{quote}}

\def\NP{{\it Nucl.Phys.} }
\def\PL{{\it Phys.Lett.} }
\def\PR{{\it Phys.Rev.} }
\def\PRL{{\it Phys.Rev.Lett.} }

\def\gappeq{\mathrel{\rlap {\raise.5ex\hbox{$>$}}
{\lower.5ex\hbox{$\sim$}}}}

\def\lappeq{\mathrel{\rlap{\raise.5ex\hbox{$<$}}
{\lower.5ex\hbox{$\sim$}}}}

\begin{document}
\begin{flushright}
{CERN-TH/98-346}
{astro-ph/9812211}
\end{flushright}
%\vspace*{5mm}
\begin{center}
{\bf PARTICLE CANDIDATES FOR DARK MATTER} \\
\vspace*{1cm} 
{\bf John Ellis}\\
\vspace{0.3cm}
Theoretical Physics Division, CERN \\
CH - 1211 Geneva 23 \\
\vspace*{3.0cm}  
{\bf Abstract} \\ \end{center}
\vspace*{5mm}
\noindent
Some particle candidates for dark matter are reviewed in the
light of recent experimental and theoretical developments.
Models for {\it massive neutrinos} are discussed in the
light of the recent atmospheric-neutrino data, and used
to motivate comments on the plausibility of
different solutions to the solar neutrino problem.
Arguments are given that the {\it lightest supersymmetric
particle} should be a neutralino $\chi$, and accelerator and
astrophysical constraints used to suggest that 
$50~{\rm GeV} \lappeq m_{\chi} \lappeq 600$~GeV. Minimizing the fine
tuning
of the gauge hierarchy favours $\Omega_{\chi} h^2 \sim 0.1$.
The possibility of {\it superheavy relic particles} is mentioned,
and candidates from string and $M$ theory are reviewed.
Finally, the possibility of non-zero {\it vacuum energy} is
discussed: its calculation is a great opportunity for a quantum
theory of gravity, and the possibility that it is time
dependent should not be forgotten.

\vspace*{5cm}
\begin{center}
{\it Invited talk presented at the Nobel Symposium}\\
{\it Haga Slott, Sweden, August 1998}
\end{center} 
%\noindent 
%\rule[.1in]{16.5cm}{.002in}

%\noindent
%5$^{*)}$ Permanent address: Kharkov Institute of Physics and Technology,

%Kharkov 310108,  Ukraine. e-mail address: dvolkov @ kfti.kharkov.ua. 
\vspace*{0.5cm}

%\begin{flushleft} CERN-TH.7226/94 \\
%April 1994
%\end{flushleft}
\vfill\eject
%\pagestyle{empty}
%\clearpage\mbox{}\clearpage

\setcounter{page}{1}
\pagestyle{plain}

%INSERT YOUR TEXT HERE
\section{Introduction}
There is a wide range of possible masses for a candidate dark matter 
particle~\cite{review}.  If it was
once in thermal equilibrium, its number density $n$ is almost independent 
of its mass, as
long as the latter is $\ll 1$~MeV.  Thus, for neutrinos with masses in 
the range discussed
in the next section, $n_\nu \sim$ constant and hence $\rho_\nu = m_\nu 
n_\nu \propto m_\nu$,
leading to
\beq
\Omega_\nu h^2 \equiv \left(\frac{\rho_{\nu}}{\rho_c}h^2\right) \simeq 
\sum_i \left(
\frac{m_{\nu_i}}{98~{\rm eV}}\right)~,
\label{one}
\eeq
where $h$ is the present Hubble expansion rate in units of 
100~kms$^{-1}$Mpc$^{-1}$ and
$\rho_c$ is the critical density.  There is a region of masses for 
neutrinos, or similar
particles, between 0(100)eV and 0(3)GeV where $\Omega_\nu > 1$.  Above 
this range, the
cosmological density $\rho_X$ of a particle $X$ may be sufficiently 
suppressed by mutual
annihilation:
\beq
\rho_X = m_Xn_X~:~n_X \propto \frac{1}{\sigma_{\rm ann}(XX \rightarrow 
...)}~,
\label{two}
\eeq
that $\Omega_X \lappeq 1$.  In the case of neutrinos or similar 
particles, annihilation is
particularly efficient when $m_\nu \sim m_{Z/2}$, leading to a local 
minimum in the relic
density with $\Omega_{\nu} \ll 1$.  Above this mass, the relic density in 
general rises again,
depending on the behaviour of the annihilation cross-section.  Thus one 
has three regions
where the relic density of a neutrino or similar particle may be of 
cosmological interest: 
$\Omega_\nu \sim 1$:
\beq
m_\nu \sim 30~{\rm eV}, ~~~\sim {\rm few~GeV}, ~~~{\rm or} \sim 100~{\rm 
GeV}~.
\label{three}
\eeq
The first of these possibilities corresponds to hot dark matter, and is 
of active
interest for neutrinos as discussed in Section~2, whilst the other two 
correspond to cold
dark matter, and are of interest for supersymmetric particles, as 
discussed in Section ~3. 
The middle region has been essentially excluded by LEP, as we discuss in 
more detail later,
whilst the third may be in the realm of the LHC.

It is of interest for our subsequent discussion to review in more
detail~\cite{Dimopoulos} the upper limit
(\ref{three}) on the mass of a cold dark matter particle.  One may write
\beq
\Omega_X = \frac{\rho_X}{\rho_c} \simeq \frac{m_Xn_X}{2\times 
10^4h^2T^2_0}~,
\label{four}
\eeq
where $T_0 \simeq 2.73K$ is the present effective temperature of the 
cosmic microwave
background radiation.  To a good approximation, the comoving number 
density has remained
essentially constant since the freeze-out temperature $T_f$ at which 
annihilation
terminated:
\beq
\frac{n_X}{T^3_0} \simeq \frac{n_X(T_f)}{T^3_f}~:
~n_X(T_f) \langle \sigma_{\rm ann}(XX)v_X \rangle = \frac{\dot a}{a} 
\simeq
\frac{T^2_f}{m_P}~,
\label{five}
\eeq
where $a$ is the cosmological scale factor and $m_P \simeq 1.2 \times 
10^{19}$~GeV is the
Planck mass.  For relic particles of interest, one typically finds that 
$m_X/T_f \sim 20$
to 30 and hence 
\beq
\Omega_X h^2 \simeq \frac{10^{-3}}{\langle\sigma_{\rm ann}(XX)v_X\rangle} 
\times \left(
\frac{1}{T_0 m_P} \right) \simeq \frac{10^{-3}}{\langle \sigma_{\rm 
ann}(XX)v_X\rangle{\rm
TeV}^2}
\label{six}
\eeq
The TeV scale emerges naturally as the geometric mean of $T_o \simeq 
2.73K$ and $m_P \sim
1.2 \times 10^{19}$~GeV!  Using the dimensional estimate $\langle 
\sigma_{\rm
ann}(XX)v_X\rangle \simeq C \cdot \alpha^2/m^2_X$, where $\alpha$ is a 
generic coupling
strength and $C$ is a model-dependent numerical coefficient, we find
\beq
m_X \simeq (16 \alpha \sqrt C) \sqrt\frac{\Omega_Xh^2}{0.25}~{\rm TeV}~,
\label{seven}
\eeq
which yields the expectation that $m_X \lappeq 1$~TeV.  We will discuss 
in Section~3 the
extent to which this argument implies that the LHC is ``guaranteed" to 
discover a cold dark
matter particle.  For the moment, we just point out that the above 
discussion assumed that
the particle was at one time in thermal equilibrium, which is not 
necessarily the case. 
Counter-examples include the axion discussed here by Turner~\cite{Turner},
and the 
superheavy relic
particles discussed here by Kolb~\cite{Kolb}, and in Section 4 of this
talk.

\section{Neutrinos}
Particle physics experiments~\cite{PDG} tell us that neutrino masses must
be much 
smaller than those
of the corresponding charged leptons and quarks:
\bea
m_{\nu_e} &\lappeq& 3.5~{\rm eV}~{\rm vs.}~m_e \simeq 0.511~{\rm 
MeV}\nonumber \\
m_{\nu_\mu} & \lappeq& 160~{\rm keV}~{\rm vs.}~m_\mu \simeq 105~{\rm 
MeV}\nonumber \\
m_{\nu_\tau} &\lappeq& 18~{\rm MeV}~{\rm vs.}~m_\tau \simeq 1.78~{\rm 
GeV}
\label{eight}
\eea
There is another difference between neutrinos and other particles, namely 
that only
left-handed neutrinos are known to exist, produced by the familiar $V-A $
charged current:
\beq
J_\mu = \bar e \gamma_\mu(1-\gamma_5)\nu_e + \bar\mu 
\gamma_\mu(1-\gamma_5)\nu_\mu +
\bar\tau \gamma_\mu(1-\gamma_5)\nu_\tau~,
\label{nine}
\eeq
whereas both quarks and charged leptons have both left- and right-handed 
states $q_{L,R},
\ell_{L,R}$.  Thus ``Dirac" masses $m^D$ coupling them are possible:
\beq
g_{h\bar ff}H_{\Delta I = 1/2, \Delta L = 0} \bar f_R f_L \Rightarrow 
m^D_f = g_{H\bar f f}\langle 0\vert H_{\Delta I = 1/2, \Delta L = 0} 
\vert 0 \rangle~,
\label{ten}
\eeq
where the quantum numbers of the Standard Model Higgs have been indicated 
explicitly.  The
following puzzles then arise.  If right-handed neutrinos $\nu_R$ exist, 
why are the $m_\nu$
(\ref{eight}) so small?  If $\nu_R$ do not exist, can neutrinos acquire 
masses:  $m_\nu
\not= 0$?

Most particle theorists believe that particles should be massless only if 
there is an exact
gauge symmetry to guarantee this, as is the case for the photon and 
gluon.  However, there
is no such candidate symmetry to guarantee $m_\nu = 0$, so most of us 
expect $m_\nu \not=
0$.  The fact that the $\nu$ have no exact gauge quantum numbers enables 
them to have
Majorana masses $m^M$, since $\bar f_R$ in (\ref{ten}) is replaced by 
$\overline{f^c_L} =
f^T_LC$, where $C$ is an antisymmetric matrix:
\beq
m^M_\nu \bar\nu^c_L \nu_L = m^M_\nu \nu^T_L C \nu_L \equiv m^M_\nu \nu_L 
\cdot \nu_L~.
\label{eleven}
\eeq
This is not possible for quarks and charged leptons, because both $q_L 
\cdot q_L$ and $\ell_l
\cdot \ell_l$ have $Q_{em} \not= 0$, whilst $q_L \cdot q_L$ also has 
non-zero colour.  Such a
Majorana neutrino mass (\ref{eleven}) would require lepton-number 
violation:  $\Delta L = 2$ and
weak isospin $\Delta I = 1$, which does contradict any sacred theoretical 
principles, but
is not provided by the Higgs fields in the Standard Model or in the 
minimal SU(5) GUT.  A
Majorana mass  term (\ref{eleven}) could in principle be provided by a 
suitable ``exotic"
Higgs field:
\beq
g_{H\nu\nu} H_{\Delta I = 1, \Delta L = 2} \nu_L \cdot \nu_L \Rightarrow 
m^M_\nu =
g_{H\nu\nu} \langle 0\vert H_{\Delta I = 1, \Delta L = 2} \vert 0 
\rangle~,
\label{twelve}
\eeq
as appears in some non-minimal GUT models such as SO(10) with a 
\underline{126} Higgs
representation.  However, there are difficulties with sich a scenario,
since one would have 
expected a $\Delta I
= 1$ Majoron particle which should have been detected via invisible $Z^0$ 
decays. 
Alternatively, one can obtain $m^M_\nu$ from a non-renormalizable 
coupling to the Standard
Model Higgs:
\beq
\frac{g_5}{M} (H_{\Delta I = 1/2} \nu_L) \cdot (H_{\Delta I = 1/2} \nu_L) 
\Rightarrow
m^M_\nu = g_5~\frac{\langle 0\vert H_{\Delta I = 1/2} \vert 0 
\rangle^2}{M}~.
\label{thirteen}
\eeq
The question then arises:  what could be the origin of the large mass 
parameter $M$?  The
most natural possibility is the exchange of a massive singlet neutrino 
field,
traditionally called a right-handed neutrino $\nu_R$, though this 
nomenclature is
somewhat anachronistic.  Given a $\nu_R$, a Dirac mass $m^D_\nu = 
g_{H\bar \nu\nu} \langle
0 \vert H_{\Delta I = 1/2} \vert 0 \rangle$ is also possible, and one 
arrives at the
famous see-saw mass matrix~\cite{seesaw}:
\beq
(\nu_L, \bar \nu_R)
\pmatrix{
0 & m^D_\nu \cr
m^D_\nu & M}
\pmatrix{ \nu_L \cr \bar\nu_R}
\label{fourteen}
\eeq
One would expect the $M^D_\nu$ entries in (\ref{fourteen}) to be of order 
$m_q$ or $m_\ell
\lappeq 0(M_W)$, so the matrix diagonalization yields the following 
approximate
eigenstates and eigenvalues:
\bea
\nu_L + 0(\frac{M_W}{M})\bar \nu_R~&:&~m_L = m^D_{\nu} M^{-1}
\left ( m^D_{\nu} \right )^{\dagger} = {\cal O}(\frac{m^2_W}{M}) 
\nonumber \\
\bar\nu_R + 0 (\frac{m_W}{M})\nu_L~&:&~m_R = M~.
\label{fifteen}
\eea
Thus there are naturally very light neutrinos if $M\gg m_W$, as would be 
expected in a
GUT with $M = {\cal O}(M_{GUT})$.

If the Dirac masses of the different neutrino generations scale like the 
corresponding
quark or charged-lepton masses, one would expect
\beq
m_{\nu_i} \simeq \frac{m^2_{q_i}~{\rm or}~m^2_{\ell_i}}{M_i}~,
\label{sixteen}
\eeq
and hence
\beq
m_{\nu_e} \ll m_{\nu_\mu} \ll m_{\nu_\tau}~, 
\label{seventeen}
\eeq
if the heavy Majorana mass matrix is diagonal, and if its eigenvalues 
$M_i$ are 
approximately the same.  As an example, putting
$m^D_\nu \sim 100$~GeV for the third generation, one finds $m_{\nu_3} 
\sim 0.1$~eV if
$M_3 \sim 10^{14}$~GeV.  Before the advent of the atmospheric neutrino 
data~\cite{SK, Soudan,MACRO}, one might also
 have expected small neutrino mixing angles, analogous to those for 
quarks, which originate
from Dirac mass matrices.
 
 This and the prejudice (\ref{seventeen}) are not necessarily supported 
by the
 atmospheric neutrino data~\cite{SK, Soudan,MACRO}, which suggest a large
mixing angle: $\sin^2
 2\theta_{\mu\tau} \gappeq 0.8$ and a mass-squared difference $\Delta 
m^2_{\rm atmo} \sim
 (10^{-2}$ to $10^{-3}$)eV$^2$.  We also recall that the solar neutrino 
data~\cite{Bahcall,Lisi} favour one
 of three possible solutions:  the large-angle MSW solution with
 $$
 \Delta m^2_{\rm solar} \sim 10^{-5}{\rm eV}^2~,~~\sin^2 2\theta \gappeq 
0.8~,
 \eqno{(18a)}
 $$
 the small-angle MSW solution with
 $$
 \Delta m^2_{\rm solar} \sim 10^{-5}{\rm eV^2}~,~~\sin^2 2\theta \sim 
10^{-2}~,
 \eqno{(18b)}
 $$
 or vacuum oscillations with
 $$
 \Delta m^2_{\rm solar} \sim 10^{-10}{\rm eV^2}~,~~\sin^2 2\theta \gappeq 
0.6~.
 \eqno{(18c)}
 $$
 \addtocounter{equation}{1}
\noindent
Which of these solutions might be favoured by post-super-Kamiokande 
models of neutrino
 masses, and how plausible is neutrino hot dark matter of cosmological 
and astrophysical
 significance, which would require $m_\nu \gappeq 1$~eV for at least one 
neutrino
 species?
 
 A first comment is that large neutrino mixing is (in retrospect) not at 
all implausible~\cite{newtheory}.
  For one thing, it is not at all necessary that $m^D_\nu \propto m_q$ or 
$m_\ell$. For
example, 
 in a specific flipped SU(5) model used earlier to discuss quark and 
lepton masses, we
 found~\cite{ELLN}
 \beq
 m^D_\nu \sim
 \pmatrix{
 {\cal O} (\eta) & {\cal O}(1) & 0 \cr
 {\cal O}(\eta) & {\cal O}(1) & 0 \cr
 0 & 0 & {\cal O}(1)}
 \label{nineteen}
 \eeq
 where $\eta$ is a small parameter, which would yield at least one large 
neutrino
 mixing angle. Moreover, there is no good reason why the heavy singlet 
Majorana
 mass matrix should be (approximately) diagonal in the same basis. For 
example, in
 models with a $U(1)$ flavour symmetry one expects matrix
elements
 \beq
 M_{ij} \propto \epsilon^{n_i+n_j} M_{GUT}
 \label{twenty}
 \eeq
  where $\epsilon$ is an expansion parameter.  If $n_i = -n_j$, one gets 
a large
  off-diagonal entry $M_{ij} = {\cal O}(1)\times M_{GUT}$, which looks at
least 
as plausible as
  having $n_i = 0$, which would be required to give a large diagonal 
entry  $M_{ii}= {\cal
O} ( M_{GUT})$. Such a large off-diagonal entry would be another source of 
large
  neutrino mixing.
  
  With so many different features contributing to the light neutrino mass 
matrix,
  there is no obvious symmetry or other reason why near-degeneracy $m_i 
 \sim m_j \gg \vert m_i - m_j\vert$ should occur. Therefore, we expect 
that there
 may be a hierarchy of masses:
 \beq
 m_{3} \sim \sqrt{\Delta m^2_{atmo}} > m_2 \sim \sqrt{\Delta 
m^2_{solar}} > m_1
 \label{twentyone}
 \eeq
 If this is indeed the case, even the heaviest light neutrino would weigh
$\lappeq$ 0.1 eV, and would not be of great astrophysical and 
cosmological interest~\cite{Tegmark}.
 
 A corollary question is the extent to which large mixing is compatible 
with a
 large neutrino mass hierarchy. To address this, we may consider a simple
two-state model~\cite{ELLN}
 \beq
 m_\nu = \pmatrix{b&d\cr d&c}
 \label{twentytwo}
 \eeq
 and diagonalize it to obtain eigenvalues $m_2$ and $m_3$ and a mixing 
angle
 $\theta_{23}$. We see in Figs. 1 and 2 that a hierarchy $m_3 = {\cal 
O}(10) m_2$ is
 quite compatible with large mixing $\sin^2\theta_{23}$ without 
fine-tuning of the 
 ratios $b/d, c/d$, for example if $b/d \sim 0.5$ and $c/d \sim 1.5$.  
Moreover, the
 light-neutrino mixing angle may be enhanced by renormalization-group 
effects
 between the GUT and electroweak scales~\cite{Tanimoto}:
 \beq
 16\pi^2 {d\over dt} ~(\sin^2\theta_{23}) = -2
 (\sin^2\theta_{23})~(\cos^2\theta_{23})\times (\lambda^2_3 - 
\lambda^2_2)~~
 {m_{33}+m_{22}\over m_{33}-m_{22}}
 \label{twentythree}
 \eeq

 \begin{figure}%fig1
 \hglue 4.5cm
 \epsfig{figure=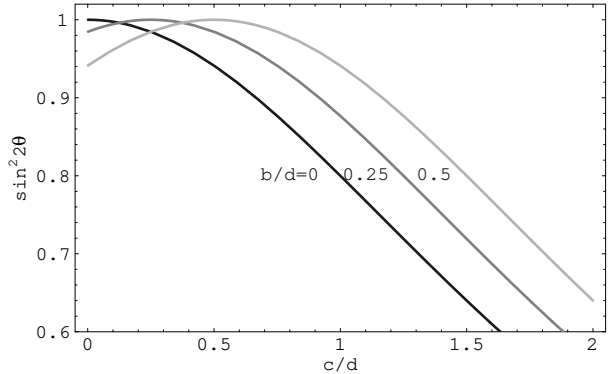,width=8cm}
 \caption[]{{\it Light-neutrino mixing in a simple $2 \times 2$ model
(\ref{twentytwo}), as a function of the 
ratios of mass-matrix elements~\cite{ELLN}.}}
 \end{figure}

\begin{figure}%fig2
 \hglue 4.5cm
 \epsfig{figure=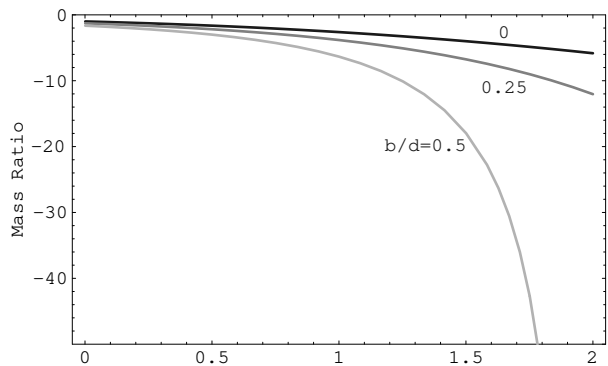,width=8cm}
 \caption[]{{\it Light-neutrino mass hierarchy in a simple $2 \times 2$
model (\ref{twentytwo}), as a function of the 
ratios of mass-matrix elements~\cite{ELLN}.}}
 \end{figure}
 
  \begin{figure}%fig3
 \hglue 4.5cm
 \epsfig{figure=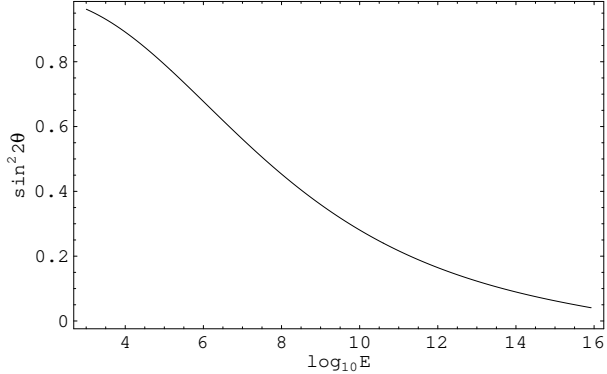,width=8cm}
\caption[]{{\it Possible renormalization-group running of the
light-neutrino mixing angle~\cite{ELLN}.}}
 \end{figure}

\noindent
 as seen in Fig. 3. The renormalization-group enhancement is particularly
important if $\lambda^2_3-\lambda^2_2$ is large, as may happen at large
 $\tan\beta$ in supersymmetric models, and/or if the diagonal etries 
$m_{33}$ and
 $m_{22}$ are almost equal. We infer that the hierarchy (\ref{twentyone}) 
is
 plausible if $\Delta m^2_{solar} \sim 10^{-5}$~eV$^2$ as in the large- 
and
 small-angle MSW solutions (18a,b). However, it 
 seems difficult to stretch the hierarchy (\ref{twentyone}) as far as 
would be
 required by the vacuum solution (18c) to the solar neutrino problem.
  
 Our present inclination is therefore to favour
 \beq
 m_{\nu_3} \sim (10^{-1}~~{\rm to}~ 10^{-1~ {1 \over 2}})~{\rm eV}~,~~ m_2 
\sim 10^{-2 ~{1 \over 2}}~{\rm eV}
 \label{twentyfour}
 \eeq 
 and the remaining question is whether the large- or small-angle MSW 
solution
 (18a,b) is to be favoured. In our specific flipped $SU(5)$
model~\cite{ELLN}, we 
found not
 only large off-diagonal entries in $m^D_\nu$ (\ref{nineteen}), but also 
large
 off-diagonal entries (\ref{twenty}) in the singlet-neutrino mass matrix:
 \beq
 M\sim \pmatrix{X&X&0\cr X&0&X \cr 0&X&X}
 \label{twentyfive}
 \eeq
 where all of the indicated non-zero entries might {\it a priori} be 
comparable in
 magnitude. We therefore find the large-angle MSW solution (18a) to be at 
least as
 plausible as the small-angle MSW solution (18b), perhaps even more so. A 
final
 comment concerns the magnitudes of the entries in (\ref{twentyfive}): we 
estimate
 their natural order of magnitude to be $0(10^{13\pm 2})$ GeV, 
corresponding to
 $m_{\nu_3} = 10^{0\pm2}$ eV, overlapping comfortably with the desired 
range
 (\ref{twentyone}).
 
 In my view, although the super-Kamiokande data~\cite{SK} make a very
strong case 
for
 atmospheric neutrino oscillations, particle physicists will not be 
completely
 convinced until they have verified the effect using a beam with 
controllable
 energy, spectrum and flavour content, as provided by an accelerator 
neutrino
 beam. One such project, the K2K experiment~\cite{KtwoK}, is to start
taking data in 
1999, and
 another, the NUMI/MINOS project~\cite{MINOS}, has also been approved. A
third 
project,
 neutrino Gran Sasso (NGS), has been studied by a joint CERN-INFN working 
group~\cite{NGS},
 and is ready for construction if funding is approved. The K2K project is 
at
 relatively low energy, insufficient to produce the $\tau$ lepton. The 
NUMI/MINOS
 and NGS projects are at higher energies, and should be able to reach 
down to
 $\Delta m^2 \sim 10^{-3}$ eV$^2$, though without much margin. 
Nevertheless, in a
 few years we may expect to know whether accelerator experiments confirm 
the
 super-Kamiokande results. However, as mentioned earlier, it currently 
seems
 unlikely that neutrinos will turn out to be an important component of 
the dark
 matter~\cite{Tegmark}.
 
 \section{The Lightest Supersymmetric Particle}
 
 The motivation for supersymmetry at an accessible energy is provided by 
the gauge
 hierarchy problem~\cite{hierarchy}, namely that of understanding why $m_W
\ll m_P$, the 
only
 candidate for a fundamental mass scale in physics. Alternatively and
 equivalently, one may ask why $G_F\sim g^2/m^2_W \gg G_N = 1/m^2_P$,
where $M_P$ is the Planck mass, expected to be the fundamental
gravitational mass scale. Or 
one may
 ask why
 the Coulomb potential inside an atom is so much
 larger than the Newton potential, which is equivalent to why $e^2 = 
{\cal O}(1) \gg
 m_pm_e/m^2_P$, where $m_{p,e}$ are the proton and electron masses.

One might think it would be sufficient to choose the bare mass
 parameters: $m_W\ll m_P$. However, one must then contend with quantum
 corrections, which are quadratically divergent:
 \beq
 \delta m^2_{H,W} = {\cal O}~\left({\alpha\over\pi}\right)~\Lambda^2
 \label{twentysix}
 \eeq
 which is much larger than $m_W$, if the cutoff $\Lambda$ representing 
the
 appearance of new physics is taken to be ${\cal O}(m_P)$. This means 
that one must
 fine-tune the bare mass parameter so that it is almost exactly cancelled 
by the
 quantum correction (\ref{twentysix}) in order to obtain a small physical 
value of
 $m_W$. This seems unnatural, and the alternative is to introduce new 
physics at
 the TeV scale, so that the correction (\ref{twentysix}) is naturally 
small.
 
 At one stage, it was proposed that this new physics might correspond to 
the Higgs
 boson being composite~\cite{technicolour}. However, calculable scenarios
of this type are
 inconsistent with the precision electroweak data from LEP and elsewhere. 
The
 alternative is to postulate approximate supersymmetry~\cite{susy}, whose
pairs of 
bosons and
 fermions produce naturally cancelling quantum corrections:
 \beq
 \delta m^2_W = {\cal O}~\left({\alpha\over\pi}\right)~\vert m^2_B - 
m^2_F\vert
 \label{twentyseven}
 \eeq
 that are naturally small: 
 $\delta m^2_W \lappeq m^2_W$ if 
\beq
\vert m^2_B - m^2_F\vert \lappeq {\rm 1 TeV}^2.
\label{twentysevenhalf}
\eeq
 There are many other possible motivations for supersymmetry, but this is 
the only
 one that gives reason to expect that it might be accessible to the 
current
 generation of accelerators and in the range (\ref{seven}) expected for a 
cold
 dark matter particle.
 
 The minimal supersymmetric extension of the Standard Model (MSSM) has 
the same
 gauge interactions as the Standard Model, and the Yukawa interactions 
are very
 similar:
 \beq
 \lambda_d QD^cH + \lambda_\ell LE^cH + \lambda_u QU^c\bar H + \mu\bar HH
 \label{twentyeight}
 \eeq
 where the capital letters denote supermultiplets with the same quantum 
numbers as
 the left-handed fermions of the Standard Model. The couplings
 $\lambda_{d,\ell,u}$ give masses to down quarks, leptons and up quarks
 respectively, via distinct Higgs fields $H$ and $\bar H$, which are 
required in
 order to cancel triangle anomalies. The new parameter in 
(\ref{nineteen}) is the
 bilinear coupling $\mu$ between these Higgs fields, that plays a 
significant
 r\^ole in the description of the lightest supersymmetric particle, as we 
see
 below. The gauge quantum numbers do not forbid the appearance of 
additional
 couplings
 \beq
 \lambda LLE^c + \lambda^\prime LQD^c + \lambda U^cD^cD^c
 \label{twentynine}
 \eeq
 but these violate lepton or baryon number, and we assume they are 
absent.
 One significant aspect of the MSSM is that the quartic scalar 
interactions are
 determined, leading to important constraints on the Higgs mass, as we 
also see
 below.
 
 Supersymmetry must be broken, since  supersymmetric partner particles do 
not have
 identical masses, and this is usually parametrized by scalar mass 
parameters
 $m^2_{0_i}\vert\phi_i\vert^2$, gaugino masses ${1\over 2} M_a\tilde
 V_a\cdot\tilde V_a$ and trilinear scalar couplings $A_{ijk}\lambda_{ijk}
 \phi_i\phi_j\phi_k$. These are commonly supposed to be inputs from some
 high-energy physics such as supergravity or string theory. It is often
 hypothesized that these inputs are universal: $m_{0_i} \equiv m_0, 
M_a\equiv
 M_{1/2}, A_{ijk}\equiv A$, but these assumptions are not strongly 
motivated by any
 fundamental theory. The physical sparticle mass parameters are then 
renormalized  in a calculable way:
 \beq
 m^2_{0_i} = m^2_0 + C_i m^2_{1/2}~,~~ M_a = \left({\alpha_a\over
 \alpha_{GUT}}\right)~~m_{1/2}
 \label{thirty}
 \eeq
where the $C_i$ are calculable coefficients~\cite{renorm}
 and MSSM phenomenology is then parametrized by $\mu, m_0, m_{1/2}, A$ 
and
 $\tan\beta$ (the ratio of Higgs v.e.v.'s).
 
 Precision electroweak data from LEP and elsewhere provide two 
qualitative
 indications in favour of supersymmetry. One is that the inferred 
magnitude of
 quantum corrections favour a relatively light Higgs boson~\cite{LEPEWWG}
 \beq
 m_h = 66^{+74}_{-39} \pm 10~{\rm GeV}
 \label{thirtyone}
 \eeq
 which is highly consistent with the value predicted in the MSSM: $m_h 
\lappeq$
 150 GeV~\cite{susymh} as a result of the constrained quartic couplings.
(On the other 
hand,
 composite Higgs models predicted an effective Higgs mass $\gappeq$ 1 TeV
and other unseen quantum corrections.)  The other indication in favour 
of
 low-energy supersymmetry is provided by measurements of the gauge 
couplings at
 LEP, that correspond to $\sin^2 \theta_W \simeq 0.231$ in agreement with 
the
 predictions of supersymmetric GUTs with sparticles weighing about 1~TeV, 
but
 in disagreement with non-supersymmetric GUTs that predict
 $\sin^2\theta_W \sim 0.21$ to 0.22~\cite{sintheta}.  Neither of these
arguments provides 
an accurate
estimate of the sparticle mass scales, however, since they are both only 
logarithmically
sensitive to $m_0$ and/or $m_{1/2}$.

The lightest supersymmetric particle (LSP) is expected to be stable in 
the MSSM, and hence
should be present in the Universe today as a cosmological relic from the 
Big Bang~\cite{EHNOS}.  This is
a consequence of a multiplicatively-conserved quantum number called $R$ 
parity, which is
related to baryon number, lepton number and spin:
\beq
R = (-1)^{3B+L+2S}
\label{thirtytwo}
\eeq
It is easy to check that $R = +1$ for all Standard Model particles and $R 
= -1$ for all
their supersymmetric partners.  The interactions (\ref{twentynine}) would 
violate $R$, but
not a Majorana neutrino mass term or the other interactions in $SU(5)$ or 
$SO(10)$ GUTs. 
There are three important consequences of $R$ conservation: (i) 
sparticles are always
produced in pairs, e.g., $pp \to \tilde{q} \tilde{g} X$, $e^+ e^- \to 
\tilde{\mu}^+
\tilde{\mu}^-$,  (ii) heavier sparticles decay into lighter sparticles, 
e.g., $\tilde{q}
\to q \tilde{g}$,
$\tilde{\mu} \to \mu \tilde{\gamma}$, and (iii) the LSP is stable because 
it has no legal
decay mode.

If such a supersymmetric relic particle had either electric charge or 
strong interactions,
it would have condensed along with ordinary baryonic matter during the 
formation of
astrophysical structures, and should be present in the Universe today in 
anomalous heavy
isotopes.  These have not been seen in studies of $H$, $He$, $Be$, $Li$, 
$O$, $C$, $Na$,
$B$ and $F$ isotopes at levels ranging from $10^{-11}$ to
$10^{-29}$~\cite{Smith}, 
which are far below
the calculated relic abundances from the Big Bang:
\beq
\frac{n_{relic}}{n_p} \; \gappeq \; 10^{-6} \; \mbox{to} \; 10^{-10}
\label{thirtythree}
\eeq
 for relics with electromagnetic or strong interactions. Except
possibly for very heavy relics, one would expect these primordial relic
particles to condense into galaxies, stars and planets, along with
ordinary bayonic material, and hence show up as an anaomalous heavy
isotope of one or more of the elements studied. There
would also be a `cosmic rain' of
such relics~\cite{Nussinov}, but this would presumably not be the dominant
source
of such particles on earth. The conflict with
(\ref{thirtythree}) is sufficiently acute that
the lightest supersymmetric
relic must presumably be electromagnetically neutral and weakly
interacting~\cite{EHNOS}.  In particular, I
believe that the possibility of a stable gluino can be excluded.  
This leaves as
scandidates for cold dark matter a sneutrino $\tilde{\nu}$ with spin 0, 
some neutralino
mixture of $\tilde{\gamma} / \tilde{H}^0 / \tilde{Z}$ with spin 1/2, and 
the gravitino
$\tilde{G}$ with spin 3/2.

LEP searches for invisible $Z^0$ decays require $m_{\tilde{\nu}} \, 
\gappeq \, 43 \;
\mbox{GeV}$~\cite{EFOS}, and searches for the interactions of relic
particles with 
nuclei then enforce 
$m_{\tilde{\nu}} \, \gappeq \,$ few  TeV~\cite{Klap}, so we exclude this
possibility 
for the
LSP.  The possibility of a gravitino $\tilde{G}$ LSP has attracted 
renewed interest
recently with the revival of gauge-mediated models of supersymmetry 
breaking~\cite{GR}, and could
constitute warm dark matter if $m_{\tilde{G}} \simeq 1 \, \mbox{keV}$.  
In this talk,
however, I concentrate on the $\tilde{\gamma} / \tilde{H}^0 /
\tilde{Z}^0$ neutralino combination $\chi$, which is the best 
supersymmetric candidate for
cold dark matter.

The neutralinos and charginos may be characterized 
at the tree level by three parameters: 
$m_{1/2}$, $\mu$
and tan$\beta$.  The lightest neutralino $\chi$ simplifies in the limit 
$m_{1/2} \to 0$
where it becomes essentially a pure photino $\tilde{\gamma}$, or $\mu \to 
0$ where it
becomes essentially a pure higgsino $\tilde{H}$.  These possibilities are 
excluded,
however, by LEP and the FNAL Tevatron collider~\cite{EFOS}.  From the
point of view 
of astrophysics and
cosmology, it is encouraging that there are generic domains of the 
remaining parameter
space where $\Omega_{\chi}h^2 \simeq 0.1$ to $1$, in particular in 
regions where $\chi$ is
approximately a $U(1)$ gaugino $\tilde{B}$, as seen in Fig.
4~\cite{EFGOS}.

 \begin{figure}%fig4
 \hglue 4.5cm
 \epsfig{figure=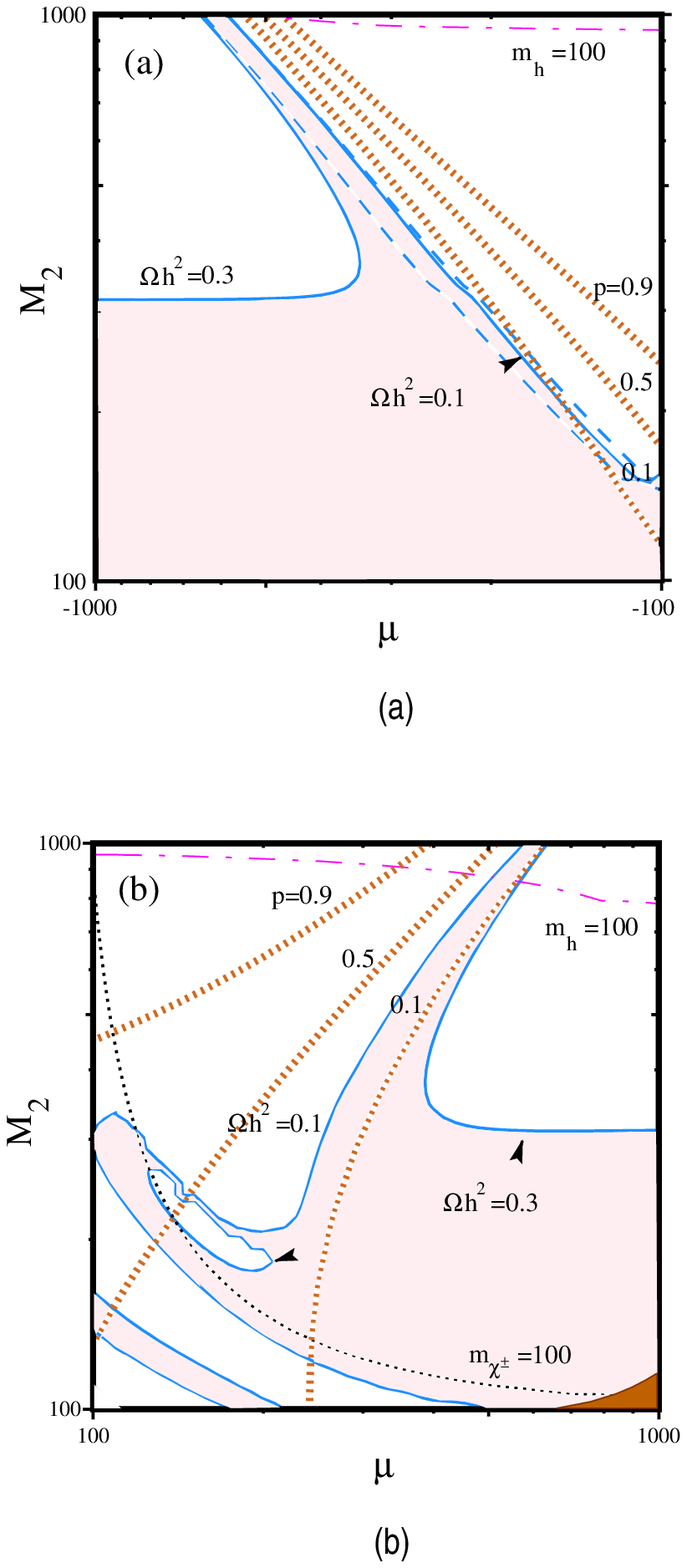,width=6cm}
\caption[]{{\it Regions of the $(\mu, M_2)$ plane in which the
supersymmetric relic density may lie within the interesting range
$0.1 \le \Omega h^2 \le 0.3$~\cite{EFGOS}.}}
 \end{figure}

 \begin{figure}%fig5
 \hglue 4.5cm
 \epsfig{figure=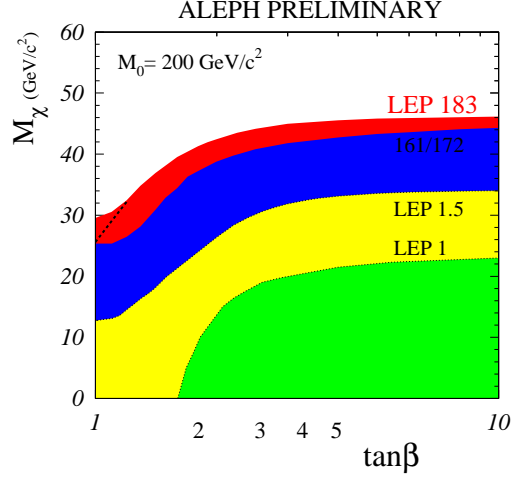,width=7.5cm}
\caption[]{{\it Experimental lower limit on the lightest neutralino mass,
inferred from unsuccessful chargino and neutralino searches at
LEP~\cite{LEPC}.}}
 \end{figure}

 \begin{figure}%fig6
 \hglue 1.5cm
 \epsfig{figure=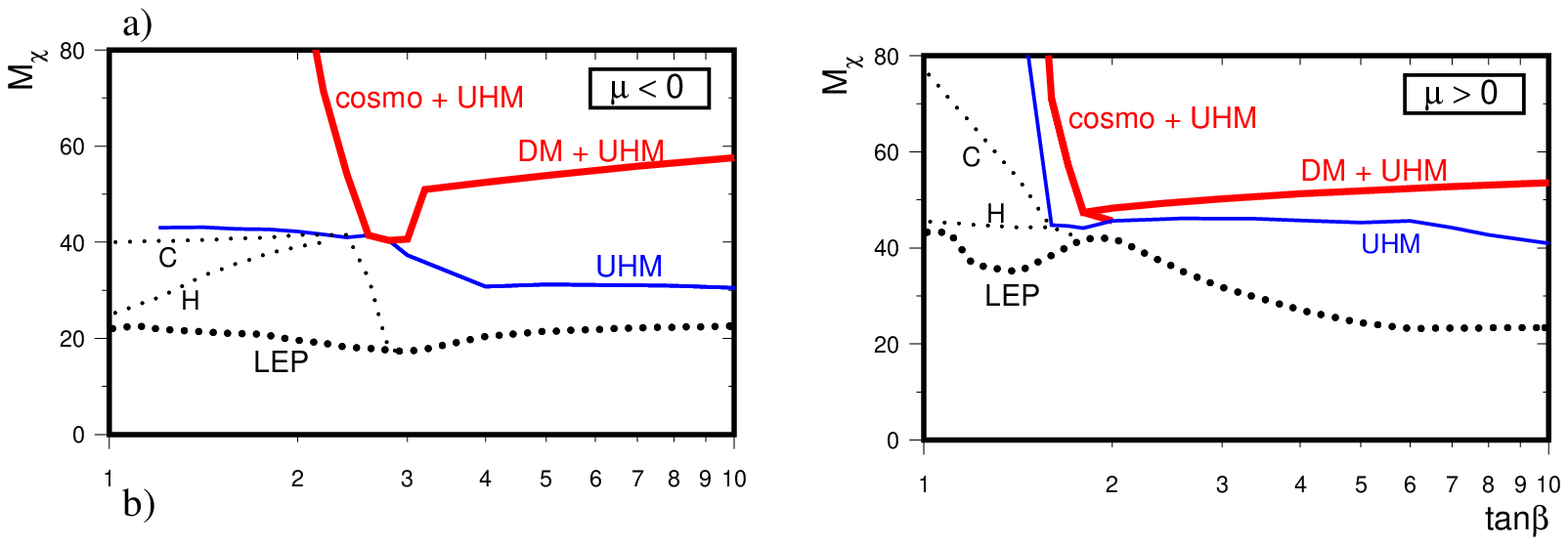,width=14cm}
 \caption[]{{\it Theoretical lower limits on the lightest neutralino
mass, obtained by using the unsuccessful Higgs searches (H), the
cosmological upper limit on the relic density (C), the assumption that
all input scalar masses are universal, including those of the Higgs
multiplets (UHM), and combining this with the cosmological upper (cosmo)
and astrophysical lower (DM) limits on the cold dark matter
density~\cite{EFOS}.}}
 \end{figure}

Purely experimental searches at LEP enforce $m_{\chi} \gappeq 30$ GeV, as 
seen in Fig. 5~\cite{LEPC}. 
This bound can be strengthened by making various theoretical assumptions, 
such as the
universality of scalar masses $m_{0_i}$, including in the Higgs sector, 
the cosmological
dark matter requirement that $\Omega_{\chi} h^2 \leq 0.3$ and the 
astrophysical preference
that $\Omega_{\chi} h^2 \geq 0.1$.  Taken together as in Fig. 6, we see 
that they enforce
\beq
m_{\chi} \gappeq 42 \; \mbox{GeV}
\label{thirtyfour}
\eeq
and LEP should eventually be able to establish or exclude $m_{\chi}$ up 
to about 50 GeV. 
As seen in Fig. 7, LEP has already explored almost all the parameter 
space available for a
Higgsino-like LSP, and this possibility will also be thoroughly explored 
by LEP~\cite{LEPC}.

 \begin{figure}%fig7
 \hglue 3cm
 \epsfig{figure=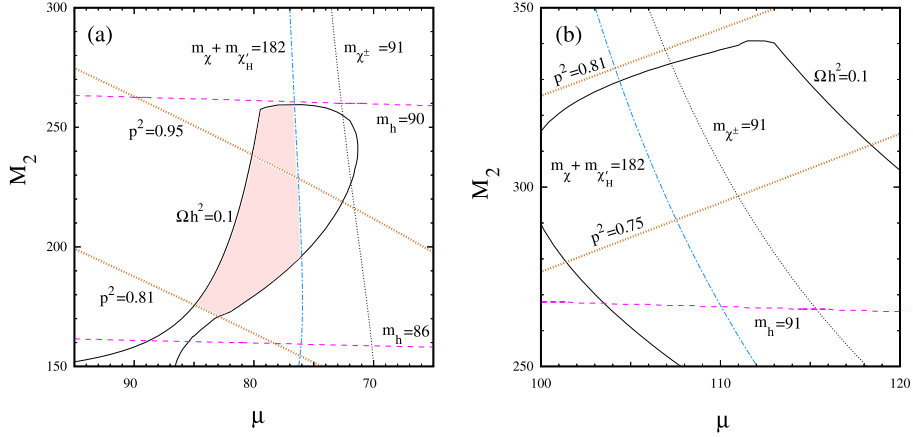,width=12cm}
\caption[]{{\it The regions of the $(\mu, M_2)$ plane where the
lightest supersymmetric particle may still be a Higgsino, taking
into account the indicated LEP constraints~\cite{EFGOS}. The
Higgsino purity is indicated by $p^2$.}}
 \end{figure}

 \begin{figure}%fig8
 \hglue 4.5cm
 \epsfig{figure=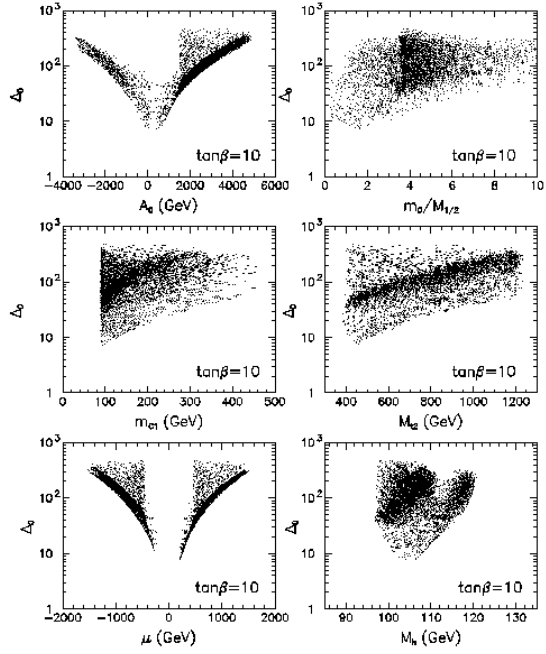,width=8cm}
\caption[]{{\it The fine-tuning price $\Delta_0$ imposed by LEP for
tan$\beta = 10$, as a function of model parameters~\cite{CEOP}.}}
 \end{figure}

 \begin{figure}%fig9
 \hglue 4cm
 \epsfig{figure=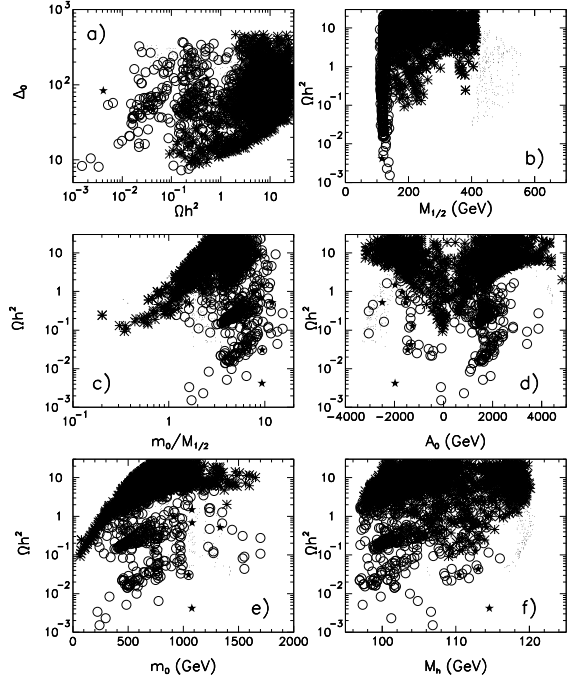,width=8cm}
\caption[]{{\it The correlation between the fine-tuning price $\Delta_0$
and the relic density $\Omega h^2$, showing dependences on model
parameters~\cite{CEOPO}.}}
 \end{figure}

 \begin{figure}%fig10
 \hglue 3cm
 \epsfig{figure=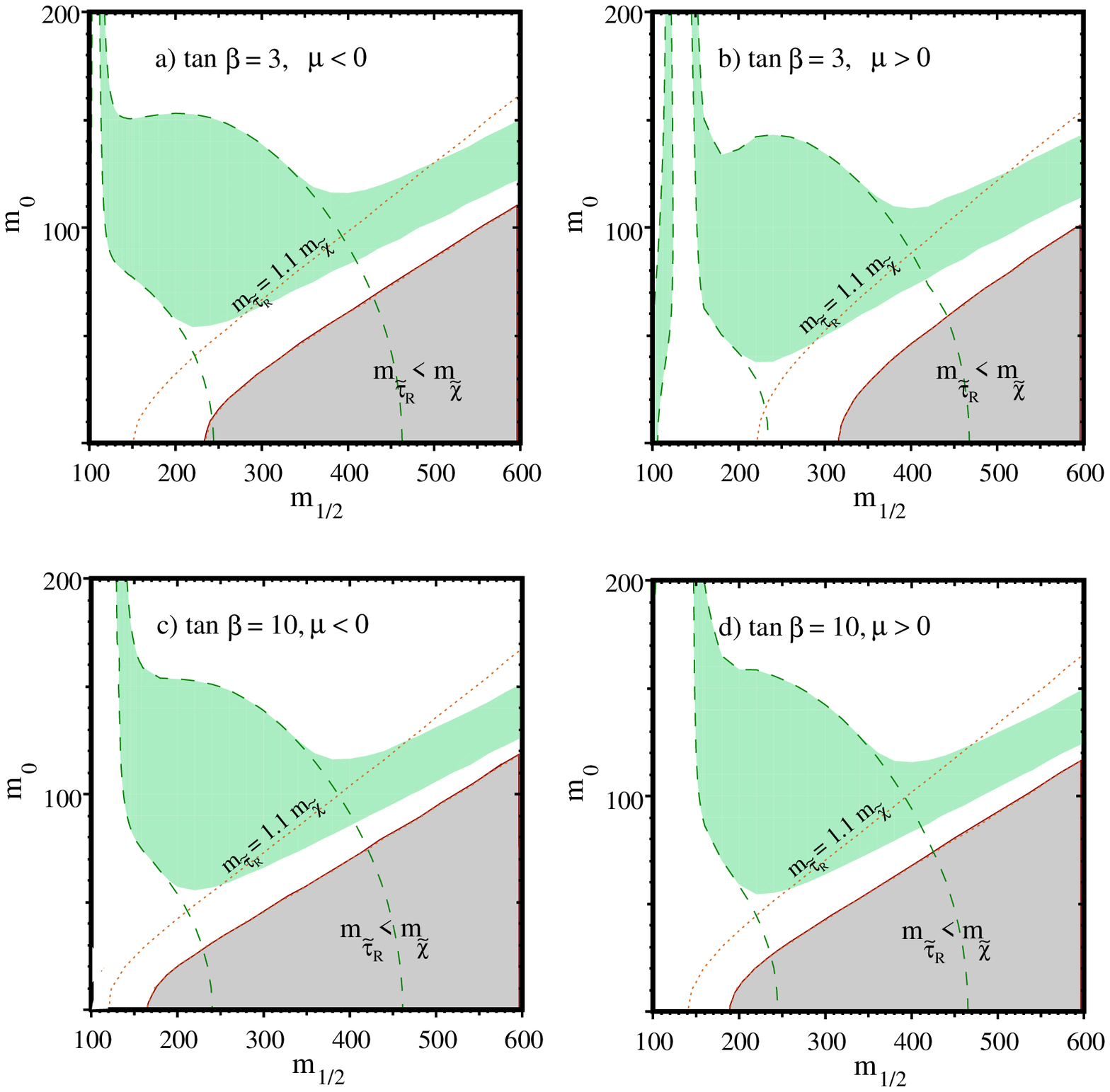,width=10cm}
 \caption[]{{\it The change in the domain of parameter space
allowed by the requirements $0.1 \le \Omega h^2 \le 0.3$ after 
(shaded region) and before (dashed lines) including $\tilde \tau$
co-annihilation~\cite{EFO}.}}
 \end{figure}

Should one be concerned that no sparticles have yet been seen by either 
LEP or the FNAL
Tevatron collider?  One way to quantify this is via the amount of 
fine-tuning of the input
parameters required to obtain the physical value of $m_W$~\cite{fine}:
\beq
\Delta_o = 
Max_{i} \;
\mid \frac{a_i}{m_W} \; \frac{\partial m_W}{\partial a_i} \mid
\label{thirtyfive}
\eeq
where $a_i$ is a generic supergravity input parameter.  As seen in Fig. 
8, the LEP
exclusions impose~\cite{CEOP}
\beq
\Delta_o \gappeq 8
\label{thirtysix}
\eeq
Although fine-tuning is a matter of taste, this is perhaps not large 
enough to be alarming,
and could in any case be reduced significantly if a suitable theoretical 
relation between
some input parameters is postulated~\cite{CEOP}.  It is interesting to
note that the 
amount of
fine-tuning $\Delta_o$ is minimized when $\Omega_{\chi}h^2 \sim 0.1$ as
preferred astrophysically, as seen in Fig. 9~\cite{CEOPO}. This means that
solving the 
gauge hierarchy
problem naturally leads to a relic neutralino density in the range of 
interest to
astrophysics and cosmology.  I am unaware of any analogous argument for 
the neutrino or the
axion.

As $m_{\chi}$ increases, the LSP annihilation cross-section decreases and 
hence its relic
number and mass density increase. How heavy could the LSP be?  Until 
recently, the limit
given was $m_{\chi} \lappeq 300$ GeV~\cite{limit}.  However, it has now
been pointed 
out that there are
regions of the MSSM parameter space where co-annihilations of the $\chi$ 
with the stau
slepton $\tilde{\tau}$ could be important, as seen in Fig. 10~\cite{EFO}.
These 
co-annihilations would
suppress $\Omega_{\chi}$, allowing a heavier neutralino mass, and we now 
find that~\cite{EFO}
\beq
m_{\chi} \lappeq \; 600 \, \mbox{GeV}
\label{thirtyseven}
\eeq
is possible.  In the past, it was thought that all the 
cosmologically-preferred region of
MSSM parameter space~\footnote{There has recently been progress in
implementing the constraints from the absence of charge and
colour-breaking minima~\cite{AF}.} could
be explored by the LHC~\cite{Abdullin}, as seen in Fig. 11, 
but it now seems
possible that there may be a delicate region close to the upper bound 
(\ref{thirtyseven}). 
This point requires further study.

 \begin{figure}%fig11
 \hglue 4cm
 \epsfig{figure=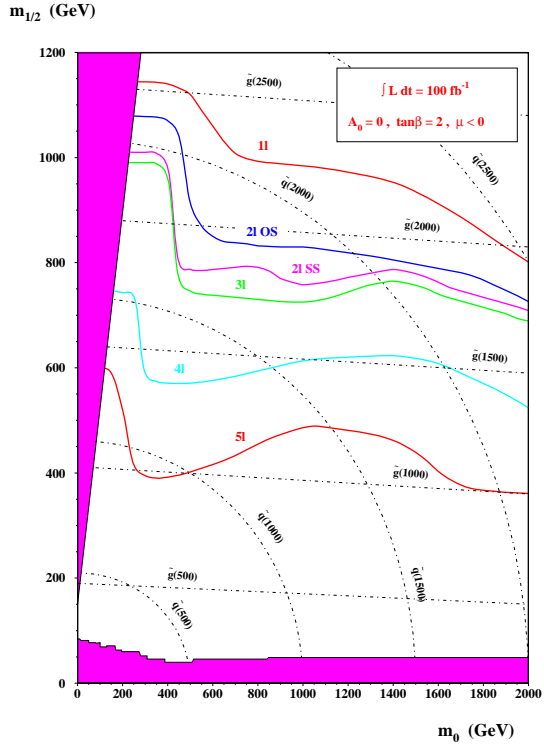,width=8cm}
\caption[]{{\it The region of the $(m_0, m_{1/2})$ plane accessible
to sparticle searches at the LHC~\cite{Abdullin}.}}
 \end{figure}

\section{Superheavy Relic Particles}

The expectation (\ref{seven}), exemplified by the MSSM range 
(\ref{thirtyseven}), was based
on the assumption that the cold dark matter particles were at one time in 
thermal
equilibrium.  As discussed here by Kolb~\cite{Kolb}, much heavier relic
particles are 
possible if one
invokes non-thermal production mechanisms.  Non-thermal decays of 
inflatons in conventional
models of cosmological inflation could yield $\Omega_{\chi} \sim 1$ for 
$m_{\chi} \sim
10^{13}$ GeV.  Preheating via the parametric resonance decay of the 
inflaton could
even yield 
$\Omega_{\chi} \sim 1$ for $m_{\chi} \sim 10^{15}$ GeV.  Other 
possibilities include a
first-order phase transition at the end of inflation, and gravitational 
relic production
induced by the rapid change in the scale factor in the early
Universe~\cite{Zilla}.  
It is therefore of
interest to look for possible experimental signatures of superheavy dark 
matter.

One such possibility is offered by ultra-high-energy cosmic rays.  Those 
coming from
distant parts of the Universe $(D \gappeq 100 Mpc)$ are expected to be 
cut off at an energy
$E \lappeq 5 \times 10^{19}$ GeV, because of the reaction $p + 
\gamma_{CMBR} \to \Delta^+$~\cite{GZK}.
However, no such Greisen-Zatsepin-Kuzmin cut-off is seen in the
data!~\cite{UHECR} The
ultra-high-energy cosmic rays must originate nearby, and should point 
back to any
point-like sources such as AGNs.  However, no such sources have been 
seen.

Could the ultra-high-energy cosmic rays be due to the decays of 
superheavy relic particles?
These should be clustered in galactic haloes (including our own), and 
hence give an
anisotropic flux~\cite{anisotropy}, but there would be no obvious point
sources. There 
have been some
reports of anisotropies in high-energy cosmic rays, but it is not clear 
whether they could
originate in superheavy relic decays.

We have analyzed~\cite{BEN} recently possible superheavy relic candidates
in string~\cite{ELN}
and/or $M$ theory.
One expects Kaluza-Klein states when six excess dimensions are 
compactified: $10 \to 4$ or
$11 \to 5$, which we call {\it hexons}.  However, these are expected to 
weigh $\gappeq
10^{16}$ GeV, which may be too heavy, and there is no particular reason 
to expect hexons to
be metastable. In $M$ theory, one expects massive states associated with 
a further
compactification: $5 \to 4$ dimensions, which we call {\it pentons}.  
Their mass could be
$\sim 10^{13}$ GeV, which would be suitable, but there is again no good 
reason to expect
them to be metastable.  We are left with bound states from the hidden 
sector of string/$M$
theory, which we call {\it cryptons}~\cite{ELN}.  These could also have
masses $\sim 
10^{13}$ GeV, and
might be metastable for much the same reason as the proton in a GUT, 
decaying via
higher-dimensional multiparticle operators.  For example, in a flipped 
$SU(5)$ model we
have a hidden-sector $SU(4) \times SO(10)$ gauge group, and the former 
factor confines
four-constituent states which we call {\it tetrons}.  Initial
studies~\cite{ELN,BEN} indicate that the
lightest of these might well have a lifetime $\gappeq 10^{17} y$, which 
would be suitable
for the decays of superheavy dark matter particles.
Detailed simulations
have been made of the spectra of particles produced 
by the
fragmentation of their decay products~\cite{Berez,BS}, and the
ultra-high-energy 
cosmic-ray data are
consistent with the decays of superheavy relics weighing $\sim 10^{12}$ 
GeV, as seen in
Fig. 12~\cite{BS}.  
 Issues to be resolved here include the roles of
supersymmetric 
particles in the
fragmentation cascades, and the relative fluxes of $\gamma, \nu$ and $p$ 
among the
ultra-high-energy cosmic rays.

 \begin{figure}%fig12
 \hglue 4cm
 \epsfig{figure=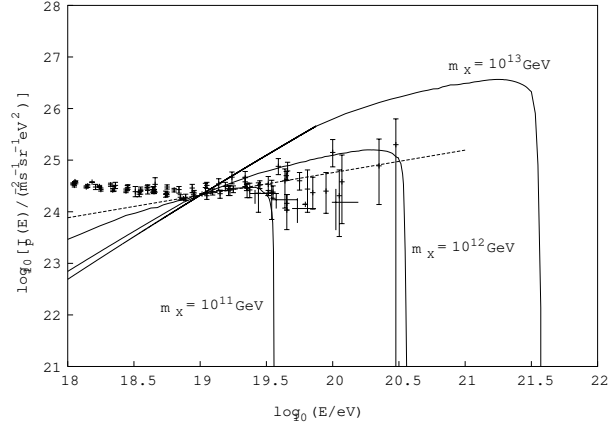,width=8cm}
\caption[]{{\it The ultra-high energy cosmic ray flux compared
with a model calculation based on the decays of superheavy relic
particles~\cite{BS}.}}
 \end{figure}

\section{Vacuum Energy}

Data on large-scale structure~\cite{BahcallN} and high-redshift
supernovae~\cite{highzSN} have recently 
converged on the
suggestion that the energy of the vacuum may be non-zero.  In my view, 
this represents a
wonderful opportunity for theoretical physics: a number to be calculated 
in the Theory of
Everything including quantum gravity.  The possibility that the vacuum 
energy may be
non-zero may even appear more natural than a zero value, since there is 
no obvious symmetry
or other reason known why it should vanish.

In the above paragraph, I have used the term {\it vacuum energy} rather 
than {\it
cosmological constant}, because it may not actually be constant.  This 
option has been
termed {\it quintessence} here by Steinhardt~\cite{Steinhardt}, who has
discussed a 
classical scalar-field
model that is not strongly motivated by the Standard Model, supersymmetry 
or GUTs, though
something similar might emerge from string theory.  I prefer to think 
that a varying vacuum
energy might emerge from a quantum theory of gravity, as the
vacuum 
relaxes towards an
asymptotical value (zero?) in an infinitely large and old Universe. We 
have recently given~\cite{EMN}
an example of one such possible effect which yields a contribution to the 
vacuum energy
that decreases as $1/t^2$.  This is compatible with the high-redshift 
supernova data, and
one may hope that these could eventually discriminate between such a 
possibility and a true
cosmological constant.

\end{document}